\documentclass[jkps,twoside,twocolumn]{revtex4}
\usepackage{graphicx}

\begin{document}
\title{Random walk and Pair-Annihilation Processes on Scale-Free Networks}
\author{Jae Dong \surname{Noh} and Sang-Woo \surname{Kim}}
\affiliation{Department of Physics, Chungnam National University,
Daejeon 305-764, Korea}

\received{\today}

\begin{abstract}
We investigate the dynamic scaling properties of stochastic particle systems
on a non-deterministic scale-free network. 
It has been known that the dynamic scaling behavior
depends on the degree distribution exponent of the underlying scale-free
network. Our study shows that it also
depends on the global structure of the underlying network.
In random walks on the tree structure scale-free network, we find that the
relaxation time follows a power-law scaling $\tau\sim N$ with the network
size $N$. And the random walker return probability decays algebraically with
the decay exponent which varies from node to node.
On the other hand, in random walks on the looped scale-free network, they
do not show the power-law scaling. We also study a pair-annihilation process
on the scale-free network with the tree and the looped structure,
respectively. We find that the particle density decays algebraically in time
both cases, but with the different exponent.

\end{abstract}


\maketitle

\section{INTRODUCTION}
Networks composed of nodes and edges have been attracting a lot of
research interest recently~\cite{review}. Unlike conventional networks such as
a periodic regular network and a random network, many real-world
networks have complex structure. Many of them belong to the class
of so-called scale-free~(SF) networks. 
Denoting the degree $k$ of a node as the
number of edges attached to it, a SF is
characterized with the power-law degree distribution
\begin{equation}
P_{deg}(k) \sim k^{-\gamma} \ .
\end{equation}
Here $\gamma$ is called the degree distribution exponent.

The power-law degree distribution implies that the SF network has an
inhomogeneous structure.
On the one hand, it is a challenging problem to characterize and
understand the organization principle of the SF network.
On the other hand, it is also interesting to study 
thermodynamic or dynamic systems on such an inhomogeneous structure.
Various physical problems have been studied.
Examples include the ferromagnetic phase transitions in the Ising 
model~\cite{Ising},
the non-equilibrium phase transition in the epidemic spreading 
model~\cite{SIS}, the random walk
process~\cite{Lahtinen01,Almaas03,Noh04a,Noh04b,Sood05,Bollt05}, 
the dynamic scaling in pair-annihilation 
process~\cite{Gallos04,Catanzaro05}, and so on.
In those studies, the $\gamma$-dependent scaling properties have been
studied. 

In the present work, we address the question how the global 
structure of the underlying SF network influences the dynamic scaling 
behavior of stochastic particle systems. 
For that purpose we study the random walk process and the pair-annihilation
process, and investigate their dynamic scaling
property. Both systems are studied on the {\em tree} structure SF~(TSF) 
networks and the {\em looped} structure SF~(LSF) networks, respectively.

To be specific, we study the stochastic systems on the
Dorogovtsev-Mendes-Samukhin~(DMS) network~\cite{DMS}, which is a
generalization of the Barab\'asi-Albert~(BA) network~\cite{BAnet}. 
It is a {\em non-deterministic} model for a growing network: Each time step, 
a new node is added, and linked with $m$ nodes which are
selected among existing nodes with the probability given by
$\Pi_i \propto  ( k_i + a )$. Here the parameter $a$ is called an
initial attractiveness. The BA model corresponds to the $a=0$ case
of the DMS network. The resulting network is scale-free and the degree
distribution exponent is given by
\begin{equation}\label{gamma}
\gamma = 3 + \frac{a}{m} \ .
\end{equation}
With the parameters $m$ and $a$, one can vary the value of the degree
exponent. At the same time, one can also generate a TSF network with 
$m=1$ or a LSF network with $m\neq 1$.
Hence, we can study the effect of the degree distribution exponent
and the global network structure systematically.

This paper is organized as follows: In Sec.~\ref{sec:2}, we present the
results on the random walk process. Section~\ref{sec:3} is devoted to the
scaling property of the pair-annihilation process. Summary will be given in
Sec.~\ref{sec:4}.

\section{Random walk process}\label{sec:2}
As a basic and fundamental stochastic process, the random walk
process~\cite{Hugh95}
on complex networks has been attracting a lot of research
interest~\cite{Lahtinen01,Almaas03,Noh04a,Noh04b,Sood05,Bollt05}.
In this work we concentrate on relaxation dynamics on SF networks.

On a network with $N$ nodes, a random walker is assigned to a starting node
denoted by $s$ at time $t=0$.
Then, at each unit time step $\Delta t = 1$, it hops to one of the
neighboring nodes selected randomly with the equal probability. 
Defining $P(i,t;s,0)$ as the
probability to find the walker at node $i$ at time $t$, 
one finds that it evolves in time as
\begin{equation}\label{master_eq}
P(i,t+1;s,0) = \sum_{j=1}^{N} \frac{A_{ij}}{k_j} P(j,t;s,0) 
\end{equation}
with the initial condition $P(i,0;s,0) = \delta_{i,s}$.
Here $\mathbf{A}= \{A_{ij}\}$ is the adjacency matrix whose elements are
$A_{ij}=1~(0)$ if two nodes $i$ and $j$ are connected~(disconnected).

The relaxation dynamics is studied with the so-called return probability
$R_s(t) \equiv P(s,t;s,0)$. For a given network,
one can solve numerically the master
equation Eq.~(\ref{master_eq}) iteratively to obtain $R_s(t)$. We then
average it over different realizations of the networks.

On a network with loops, the probability distribution converges to
the stationary one $P_{stat.}(i) = k_i / (\sum_j k_j
)$ in the $t\rightarrow\infty$ limit~\cite{Noh04a}. Hence, the
return probability converges to $R_s(t=\infty) = k_s / (\sum_j
k_j)$. On the other hand, on a tree network, the probability
distribution is oscillating in time, so is the return probability.
The random walker cannot return to a starting node in odd time
steps, which means that $R_s(t)=0$ at odd $t$. In this case, we only
measure the return probability at even time steps, which converges
to stationary value $2 k_s / (\sum_j k_j )$.

In a SF network, the node degree is distributed so broadly that one can
define an exponent $q_s$ for each node $s$ describing the degree scaling 
with the network size: 
\begin{equation}
k_s \sim N^{q_s} \ .
\end{equation}
For instance, in the DMS network with the degree distribution exponent
$\gamma$, a {\em peripheral node} with the minimum degree has $q=0$,
while the {\em hub} with the maximum degree has
$q=1/(\gamma-1)$. Then, the return probability in the stationary state
scales with the network size as
\begin{equation}\label{R_s_scaling}
R_s (t\rightarrow\infty) \sim  N^{-(1-q_s)} \ .
\end{equation}

\begin{figure}[t!]
\includegraphics*[width=7cm]{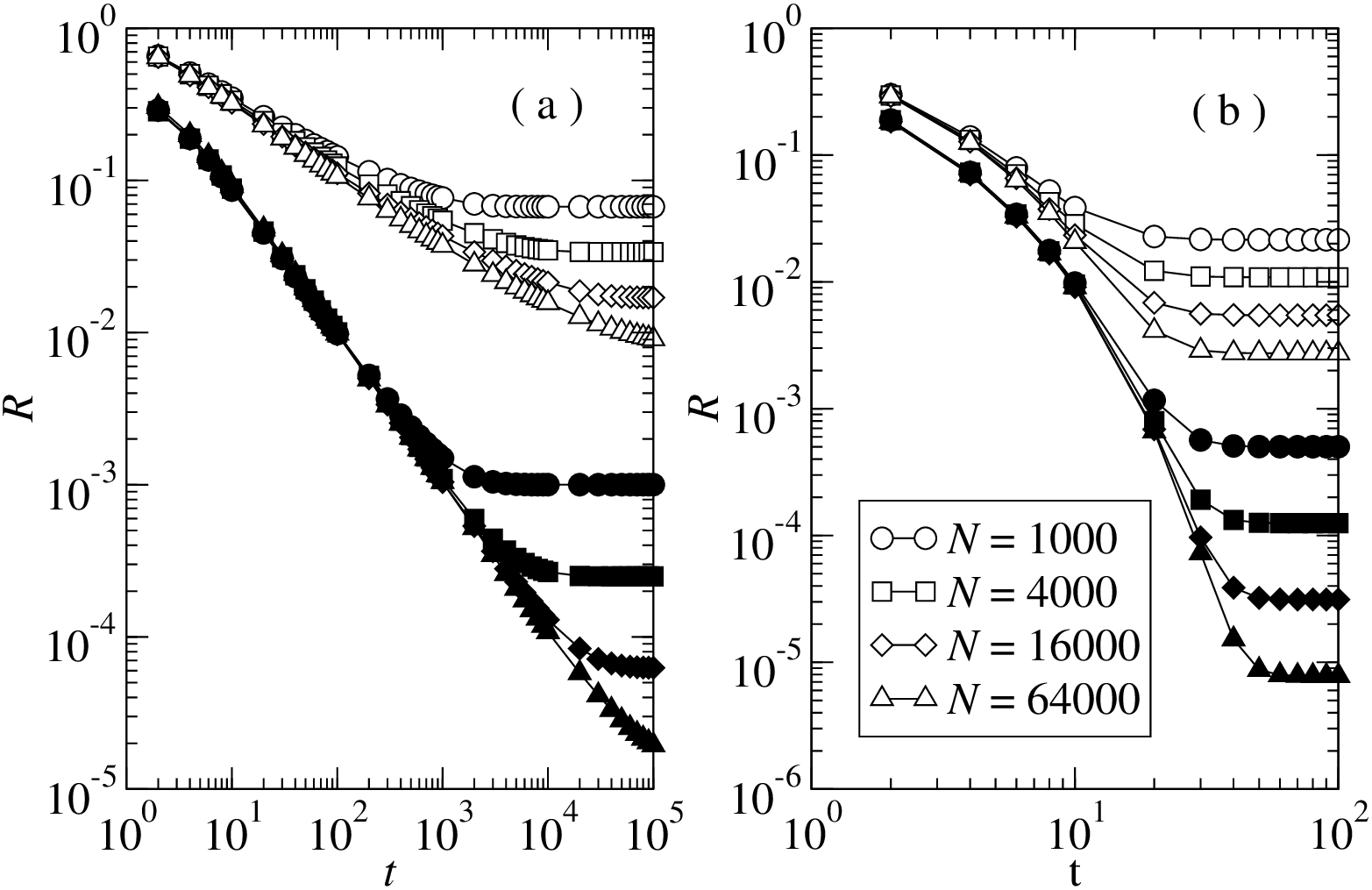}
\caption{The return probability $R_H$~(open symbols) and $R_P$~(filled
symbols) in the DMS networks with $\gamma=3$ and with $m=1$
in (a) and $m=2$ in (b).}
\label{fig1}
\end{figure}

Due to the broad degree distribution, the
return probability $R_s(t)$ may have a different scaling property at
different starting node $s$ with different values of $q_s$.
So, we measured the return probability $R_H(t)$ for the hub and $R_P(t)$ 
for a peripheral node on the DMS networks.
We present the numerical data for the return probability $R_H$ and $R_P$
in Fig.~\ref{fig1}. 
We compare the data obtained from the TSF networks and the LSF networks.
The data show that the return probability behaves distinctly depending on
the global structure of the network, which will be discussed in detail
in the following.

\subsection{Random walks on tree networks}
\begin{figure}[t!]
\includegraphics*[width=7cm]{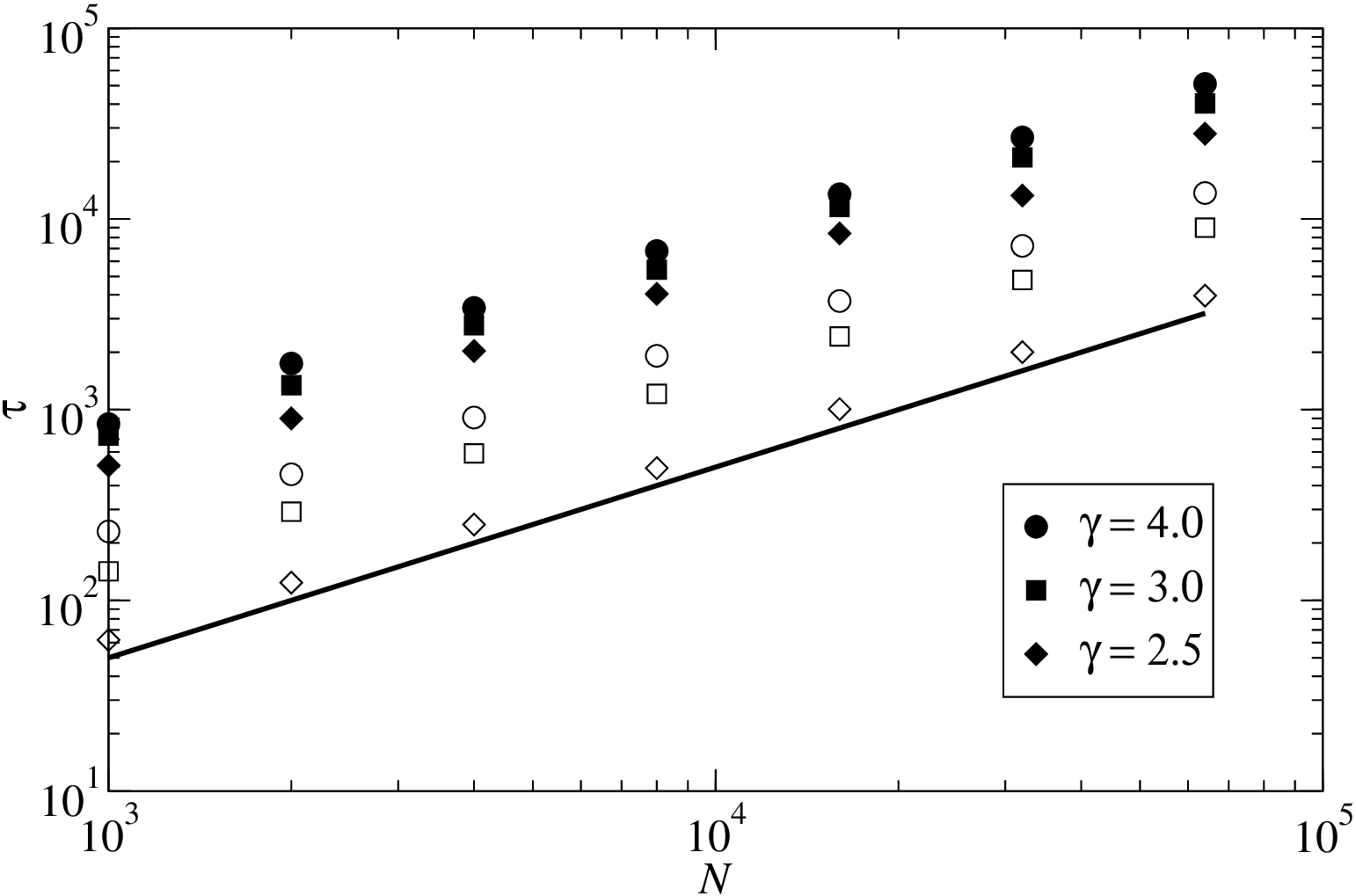}
\caption{Relaxation time $\tau$ for $R_H(t)$~(open symbols) and
$R_P(t)$~(filled symbols) on the tree structure SF networks. The solid line has
the slope 1.}
\label{fig2}
\end{figure}

We perform the quantitative analysis of the scaling property in 
the TSF. From Fig.~\ref{fig1}, one finds that 
the return probability relaxes much slower in the TSF network with $m=1$. 
We estimated the relaxation time $\tau$ using the condition $R_s(t=\tau)
= c R_s(t=\infty)$ with a constant $c=2$. The relaxation time 
is plotted as a function of the network 
size $N$ in Fig.~\ref{fig2}.
It shows that the relaxation time estimated from $R_H$
and $R_P$ follows a power-law scaling 
\begin{equation}\label{tau_N}
\tau \sim N^z
\end{equation}
with the same dynamic exponent
\begin{equation}
z=1.0 \ .
\end{equation}

We provide a theoretical argument for the numerical result from 
the analysis of the mean first passage time~(MFPT).
The MFPT problem has been studied on complex networks
recently~\cite{Noh04a,Noh04b,Bollt05}, and some rigorous results are
known~\cite{Noh04a,Noh04b}.
Consider the MFPT, denoted by $T_{j,i}$, from an arbitrary node 
$i$~(degree $k_i$) to one of its neighboring node $j$.
Let $\{n_1,n_2,\cdots,n_{k_i}=j\}$ denote the neighbors of the node $i$.
Then, following Ref.~\cite{Noh04b}, the MFPT on {\em tree} networks 
satisfies the recursion relation
\begin{equation}
T_{j,i} = k_i  + \sum_{l\neq k_i} T_{i, n_l} \ .
\end{equation}
Applying the recursion relation repeatedly until one arrives at dangling
ends, one can find the explicit solution for $T_{j,i}$.
Without the link between $i$ and $j$, the tree network would be decomposed 
into two parts. Denoting the
number of nodes in the $i$ side by $N_i$, the MFPT is simply
given by 
\begin{equation}\label{T_ji}
T_{j,i} = 2N_i -1  \ .
\end{equation} 
Hence, for a typical adjacent nodes $i$ and $j$, one has that 
$T_{j,i} \sim N^1$. 
For non-adjacent nodes $i$ and $j$, the MFPT is given by the sum of 
the MFPT's given by Eq.~(\ref{T_ji}) along the path between them. 
Hence, the MFPT between
a typical node pair is given by $T\sim D_N N$, where $D_N$ is the mean
diameter of the networks. For SF networks, the mean diameter scales at
most logarithmically with the network size~\cite{Cohen03}. Therefore, we
conclude that the relaxation time follows the power-law scaling $\tau \sim
N^z$ with the dynamic exponent $z=1$.

The power-law scaling of the relaxation time suggests that the return
probability decays algebraically as 
\begin{equation}
R_s(t) \sim t^{-\delta_s} 
\end{equation} 
with the decay exponent $\delta_s$.
For a node $s$ with the degree scaling $k_s \sim N^{q_s}$,
the decay exponent can be deduced from the finite-size-scaling ansatz
\begin{equation}\label{fss}
R_s(t) = N^{-(1-q_s)} f(t/N^z)
\end{equation}
with the dynamic exponent $z=1$.
For large $t\gg N^z$, the return probability should converge to the 
stationary value $R_s(t=\infty) \sim 
N^{-(1-q_s)}$~(see Eq.~(\ref{R_s_scaling})). 
It requires that the scaling function should behave as $f(x\gg 1)
=\mbox{constant}$. The power-law scaling for $t\ll N^z$ requires that
the scaling function should behave as $f(x\ll 1) \sim x^{-\delta_s}$, which
yields that $R_s \sim t^{-\delta_s} N^{\delta_s z - (1-q_s)}$. 
Therefore, the finite-size-scaling ansatz
predicts that the decay exponent is given by
\begin{equation}\label{delta_s}
\delta_s = 1-q_s \ .
\end{equation}

\begin{figure}[t!]
\includegraphics*[width=7cm]{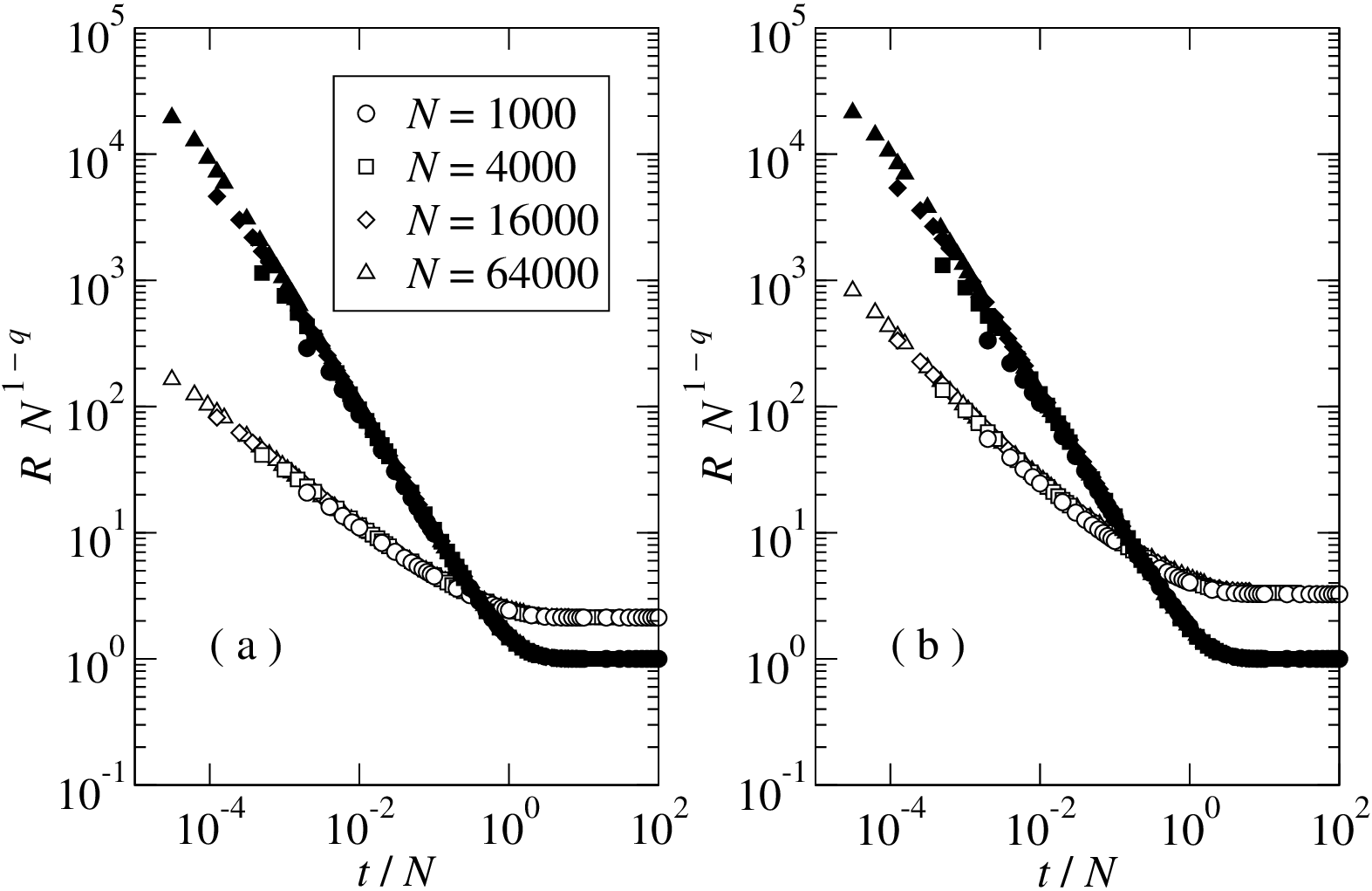}
\caption{Scaling analysis of $R_H$~(open symbols) and $R_P$~(filled symbols) 
in the TSF networks
with $\gamma=3$~(a) and $\gamma=4$~(b). }
\label{fig3}
\end{figure}

It is interesting to note that the decay exponent varies with the degree
scaling exponent $q_s$ of the starting node $s$.
We confirm the scaling behavior with the scaling plot of 
$R_s N^{1-q_s}$ versus $t/N$ for the hub and the peripheral node
in Fig.~\ref{fig3}.
For $\gamma=3$, $q_H=1/(\gamma-1)=1/2$ for the hub and $q_P=0$
for the peripheral node, which yields that $\delta_H = 1/2$ and $\delta_P = 1$. 
Similarly, for $\gamma=4$, one expects that 
$\delta_H = 2/3$ and $\delta_P = 1$.
One finds that all data from different network sizes collapse
very well in the scaling plot, which supports the result in
Eq.~(\ref{delta_s}).

It is also interesting that the return probability decays as $R_s(t)\sim
t^{-\delta_s}$ with the exponent $\delta_s \leq 1$ at all nodes.
It indicates that the random walks are recurrent in the $N\rightarrow\infty$
limit~\cite{Hugh95}.

\subsection{Random walks on looped networks}
From Fig.~\ref{fig1}, one can see that the return probability 
decays much faster in the LSF networks.
The downward curvature in the log-log plot implies that the 
decay is faster than a power-law decay.
The relaxation time measurement also indicates a faster decay.
We found that the relaxation time, measured using the condition that 
$R_s(\tau) = 2 R_s(\infty)$, grows at most logarithmically
with the network size $N$. This is in contrast to the power-law
growth in the TSF networks.
In this subsection, we address a question how the return probability decays
in time in looped SF networks. 

In the random network, it is known that the
return probability follows a stretched exponential decay as
\begin{equation}\label{stretched_exp}
R_s(t) - R_s (\infty) \sim e^{-a t^\theta}
\end{equation}
with a constant $a$ and the exponent $\theta = 1/3$~\cite{Bray88}.
It was reported that the return probability in the small-world network 
follows the stretched exponential decay, too~\cite{Jespersen00}.
It might suggest that the looped SF network follow the stretched 
exponential decay, too. 

\begin{figure}
\includegraphics*[width=7cm]{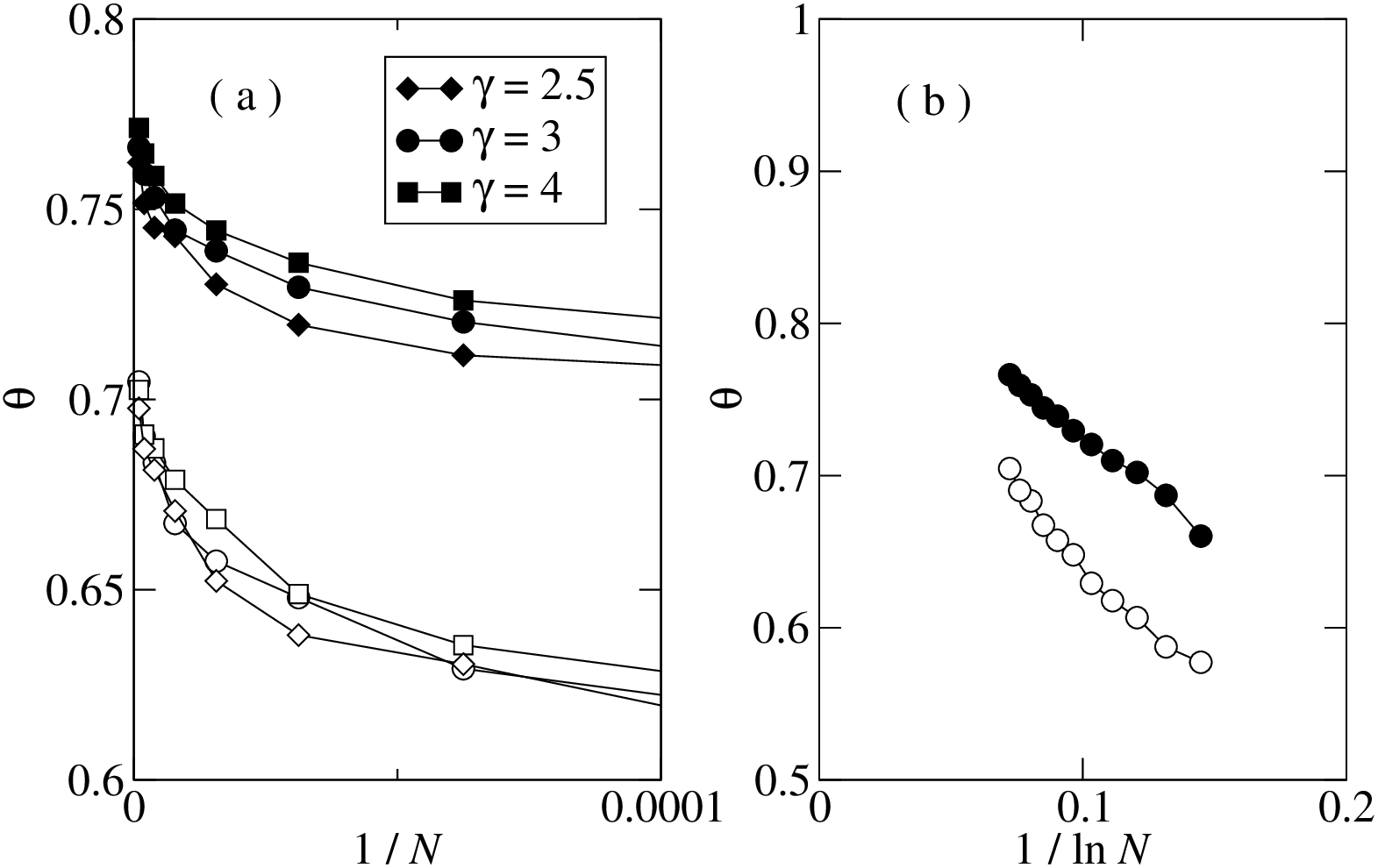}
\caption{Plots of $\theta$ obtained from $R_P(t)$ on the DMS networks with
$\gamma=2.5, 3, 4$ and with $m=2$~(open symbols) and $m=4$~(filled symbols).
$\theta$ is plotted against $1/N$ in (a) and $1/\ln N$ in (b).}
\label{fig4}
\end{figure}

On the DMS networks with $m\geq 2$, we measured $R_s(t)$ numerically
and fitted the data 
to the form in Eq.~(\ref{stretched_exp}) to estimate $\theta$. 
If the return probability follows an exponential decay, one would have
$\theta=1$. On the other hand, one would have $\theta<1$, if it follows the
stretched exponential decay.
In Fig.~\ref{fig4}, we present the data for $\theta(N)$ for $R_P(t)$ 
obtained on the DMS networks with various values of 
$\gamma=2.5, 3, 4$ and $m=2,4$ up to sizes $N\simeq 10^6$. 
Similar behaviors are observed in $\theta(N)$ for $R_H(t)$.
At small values of $N$, the exponent seems
to depend on $m$ and to be smaller than $1$. 
One may interpret that as an evidence of the stretched exponential decay 
with a non-universal exponent $\theta$. 

However, we observe that there is a very strong finite size effect. 
The plot of $\theta$ versus $1/\ln N$ in Fig.~\ref{fig4}~(b) shows that
the finite size effect is significant even for $N\simeq 10^6$.
With this strong finite size effect, one can not exclude the possibility of
the exponential decay with $\theta=1$.
We suggest that an analytic approach be
necessary to conclude whether the return probability follows the exponential
or the stretched exponential decay.

Before closing this section, we remark on the previous studies on the random
walks on {\em deterministic} SF networks~\cite{Noh04b,Bollt05}. Unlike the
non-deterministic LSF networks studied in this work, the deterministic LSF
networks~\cite{Dorogovtsev02,Jung02,Ravasz03} behave similarly as the 
TSF networks. For example, in the
hierarchical network~\cite{Ravasz03} which has a looped structure, 
the relaxation time scales algebraically as $\tau \sim N$~\cite{Noh04b}.
The reason why the hierarchical network behaves as TSF networks is clear. 
The network has a symmetry, due to which the random walks on it can be
mapped to the walks on TSF networks~\cite{Noh04b}. Therefore, our general
conclusion should not be applied to the deterministic LSF networks with high
symmetry.

\section{Pair-annihilation process}\label{sec:3}

The pair-annihilation process is a diffusion-limited reaction-diffusion
process. In this process, each node in a given network may be occupied by a
particle~(denoted by $A$) or empty~(denoted by $\emptyset$). 
The particles perform random walks on the network, and annihilate
pairwise whenever they meet at a same node~($A+A\rightarrow
\emptyset$).

The pair-annihilation process on SF networks was studied numerically by
Gallos and Argyrakis~\cite{Gallos04}. They found that the particle density
decays algebraically $\rho(t) \sim t^{-\alpha}$
with the $\gamma$-dependent decay exponent $\alpha=\alpha(\gamma)\geq 1$. 
This is contrasted with the $d$-dimensional periodic lattice case where
$\alpha=\alpha(d)\leq 1$. Namely, the particle density decays faster in SF
networks.

Later on, Catanzaro {\em et al.}~\cite{Catanzaro05} 
developed a mean field theory for the pair-annihilation process,
which will be reviewed briefly hereafter.
Let us define $\rho_k$ as the average particle density at nodes with 
degree $k$. It is related to the total density through the relation
$\rho(t) = \sum_k P_{deg}(k) \rho_k (t)$. 
In a mean field level, one can write down the rate equation for $\rho_k$ as
\begin{equation}\label{rho_k}
\frac{d\rho_k(t)}{dt} = -\rho_k(t) + \frac{k}{\langle k \rangle} [ 1- 2
\rho_k(t) ] \rho(t) \ ,
\end{equation}
where $\langle k\rangle$ is the mean degree.
Multiplying $P_{deg}(k)$ and summing over $k$, one finds that
\begin{equation}\label{rho}
\frac{d\rho(t)}{dt} = -2 \rho(t) \left[ \frac{1}{\langle k\rangle} \sum_k k
P_{deg}(k) \rho_k(t) \right] \ .
\end{equation}
Then, Catanzaro {\em et al.} made a quasistatic approximation 
neglecting the time derivative in Eq.~(\ref{rho_k}). The approximation
assumes that the particles adjust themselves so efficiently 
that their distribution $\rho_k$ remains close to a stationary one at
a given value of $\rho(t)$. It leads to the relation
\begin{equation}\label{rho_k_assume}
\rho_k(t) = \frac{ k \rho(t) / \langle k \rangle}
                 { 1 + 2  k \rho(t) / \langle k \rangle} \ .
\end{equation}
Substituting it in Eq.~(\ref{rho}) and solving the resulting equation,
Catanzaro {\em et al.} obtained that the density decay follows a power law
in the $N\rightarrow\infty$ limit as
\begin{equation}\label{rho_decay}
\rho(t) \sim t^{-\alpha} \left( \ln t \right)^{-\beta} \ .
\end{equation}
The decay exponent $\alpha$ is given by
\begin{equation}\label{alpha}
\alpha(\gamma) = \left\{
\begin{array}{lll}
 1/(\gamma-2) &,& 2 < \gamma < 3 \\
 1            &,& 3 \leq \gamma
\end{array} \right.
\end{equation}
and the exponent for the logarithmic correction is given by 
$\beta = 1$ for $\gamma=3$ and $\beta=0$ for $\gamma\neq 3$.
This result is qualitatively consistent with the simulation results
of Refs.~\cite{Gallos04,Catanzaro05}.

In the previous works~\cite{Gallos04,Catanzaro05}, 
only the $\gamma$-dependent scaling behaviors have been studied.
However, the study on the random walks in Sec.~\ref{sec:2} suggests that the
scaling behavior may also depend on the global structure of the underlying
SF network. In this section, we present the results of our numerical works on 
the pair-annihilation process on the DMS networks with $m=1$~(tree
structure) and $m\neq 1$~(looped structure).  Comparing the two different
cases, we will show that the global structure does also matter for the scaling
behavior of the pair-annihilation process.

\begin{figure}
\includegraphics*[width=7cm]{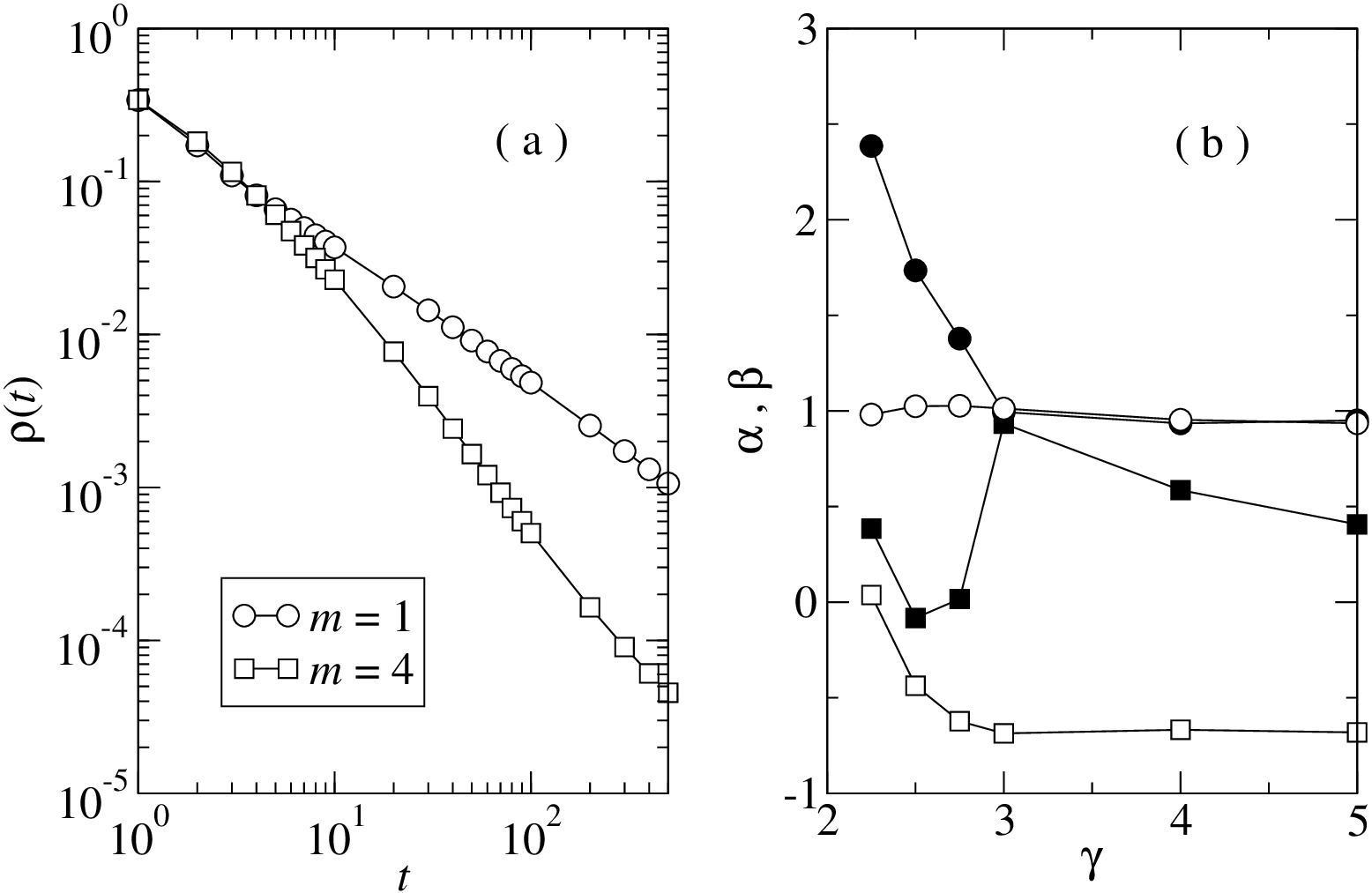}
\caption{(a) Density decay in the DMS networks with $m=1$ and $m=4$.
The degree distribution exponent is $\gamma=2.5$. 
(b) Density decay exponent $\alpha$~(circles) and $\beta$~(squares)
defined in Eq.~(\ref{rho_decay})
for the DMS networks with $m=1$~(open symbols) and $m=4$~(filled symbols).}
\label{fig5}
\end{figure}
We have performed the Monte Carlo simulations on the DMS networks of size
$N=1024000$ with various values of $\gamma$ and $m$. 
The fully-occupied state is taken as the initial configuration.
In Fig.~\ref{fig5}~(a), the numerical data
from the TSF networks~($m=1$)  and the LSF networks~($m=4$)
are compared. The data show that the particle density decay follows the
power law in both cases but with a different exponent.

We estimate the decay exponent by fitting the data to the form in
Eq.~(\ref{rho_decay}), and plot $\alpha$ and $\beta$ as a function of
$\gamma$ in Fig.~\ref{fig5}~(b). At $m=4$, our result is qualitatively 
consistent with the previous works;
$\alpha\simeq 1$ for $\gamma\geq 1$, and $\alpha$ varies
with $\gamma$ for $\gamma<3$. And the logarithmic correction is prominent at
$\gamma=3$. 
However, at $m=1$, we obtain a completely different result; $\alpha\simeq
1.0$ at all values of $\gamma$ and the logarithmic correction is present at
all values of $\gamma$.
These numerical results show that the global structure of the underling SF
networks affects the scaling behavior of the pair-annihilation process.

We speculate the origin for the different scaling behaviors. 
On the LSF networks, the scaling behavior seems to be 
consistent with the analytic mean field result in 
Eq.~(\ref{alpha}).
In the analytic approach one adopts the quasistatic
approximation~\cite{Catanzaro05} leading to the 
particle distribution given by Eq.~(\ref{rho_k_assume}).
It can be rewritten as 
\begin{equation}\label{crossover}
\rho_k(t) \simeq \left\{
\begin{array}{lll}
k \rho(t) / \langle k \rangle &,&  \mbox{for\quad} k \ll \langle k\rangle /
\rho(t) \ , \\
1 / 2 & ,& \mbox{for\quad} k \gg \langle k\rangle / \rho(t)  \ .
\end{array}\right.
\end{equation}
Note that $\rho_k \propto k$ for small $k$.

\begin{figure}[t!]
\includegraphics*[width=7cm]{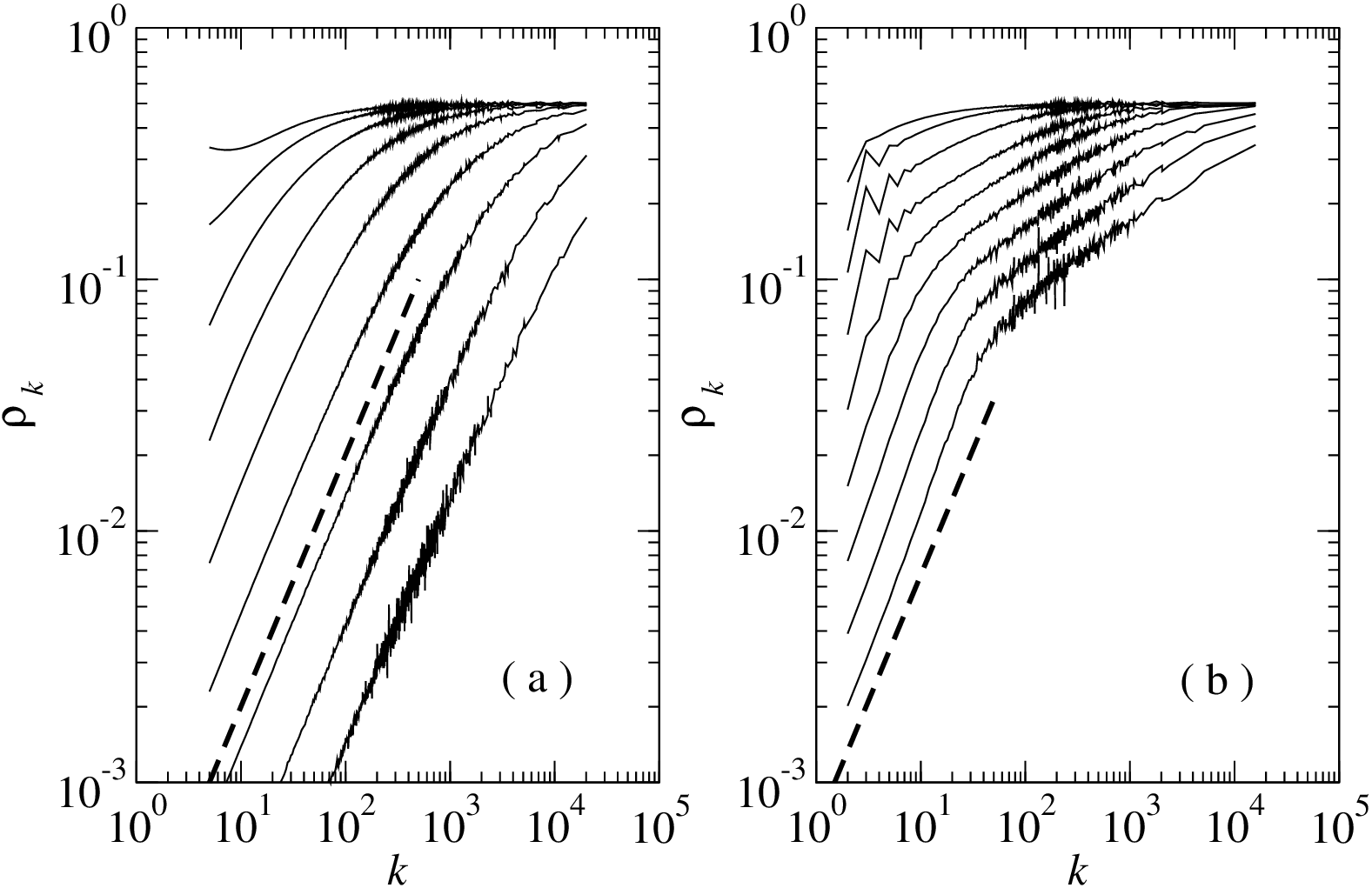}
\caption{$\rho_k$ vs. $k$ at different time steps
$t=$$2^0$~(top),$\cdots$,$2^8$~(bottom). 
The data are taken from the DMS network with $m=4$~(a) and $m=1$~(b).
The network size is $N=1024000$ and the degree distribution exponent is
$\gamma=2.5$ in both plots. The dashed lines have the slope 1.}
\label{fig6}
\end{figure}

We investigate the particle distribution numerically.
In Fig.~\ref{fig6}~(a), we plot $\rho_k(t)$ against $k$ for the DMS network
with the looped structure~($m=4$). We find that the particle distribution
$\rho_k(t)$ is fully consistent with Eq.~(\ref{crossover}).
This supports that the analytic approach is appropriate on the 
LSF networks.

However, the particle distribution deviates from Eq.~(\ref{crossover})
on the TSF network. Figure~\ref{fig6}~(b) shows
that there are three different regimes: $\rho_k \sim k$ for $k < k_1$, 
$\rho_k \sim k^\eta$ for $k_1 < k < k_2$, and $\rho_k \simeq 1/2$ for $k_2 <
k$. The scaling exponent $\eta$ in the intermediate regime is found to be
less than 1 and  to vary with $\gamma$. 
This feature is inconsistent with the quasistatic
approximation.

The quasistatic approximation assumes that particles rearrange 
themselves quickly upon the change in the total particle density via
diffusion. It requires that the diffusion should be a fast process.
In the previous section, we have shown that the diffusion is a slow 
process in the LSF networks.
This explains why the quasistatic approximation is invalid in the 
TSF network whereas it is valid in the LSF network.

\section{Summary}\label{sec:4}
In summary, we have investigated the scaling properties of the random walk
and the pair-annihilation processes on non-deterministic 
SF networks with the tree structure
and the looped structure, respectively. 
In the random walk process on TSF networks, we find that
the relaxation time scales as $\tau \sim N$ with the network size $N$ 
and that the return probability decay follows the power law with the
node-dependent exponent. The LSF network does not display
the power-law scalings. 
In the pair-annihilation process, we find that the exponent describing the
particle density decay is different in TSF and LSF networks.
Our results show that the global structure of the SF network, as well as 
the degree distribution exponent, is the important ingredient 
in understanding the dynamic scaling behaviors. 

\begin{acknowledgments}
This work was supported by Chungnam National University through the
Internal Research Grant in 2004.
\end{acknowledgments}

\end{document}